\documentclass[twocolumn,showpacs,preprintnumbers,amsmath,amssymb]{revtex4}

\usepackage{graphicx}% Include figure files
\usepackage{dcolumn}% Align table columns on decimal point
\usepackage{bm}% bold math

\newcommand{\LCMO}{La$_{0.7}$Ca$_{0.3}$MnO$_{3}$}
\newcommand{\LCMOx}{La$_{1-x}$Ca$_{x}$MnO$_{3}$}
\newcommand{\TMI}{$T_{\rm MI}$}

\newcommand{\Eb}{$E_{\rm b}$}

\begin{document}

\title{Polaronic signature in the metallic phase of La$_{0.7}$Ca$_{0.3}$MnO$_{3}$ films detected by scanning tunneling spectroscopy}

\author{S. Seiro}%
 \email{silvia.seiro@physics.unige.ch}
\author{Y. Fasano}
\author{I. Maggio-Aprile}
\author{E. Koller}
\author{O. Kuffer}
\author{\O. Fischer}

\affiliation{%
D\'epartement de Physique de la Mati\`ere Condens\' ee,
Universit\'e de Gen\`eve, \\24, Quai Ernest-Ansermet, 1211 Geneva,
Switzerland
}%

\date{\today}

\begin{abstract}

In this work we map tunnel conductance curves with nanometric
spatial resolution, tracking polaronic quasiparticle excitations
when cooling across the insulator-to-metal transition in
La$_{0.7}$Ca$_{0.3}$MnO$_{3}$ films. In the insulating phase the
spectral signature of polarons, a depletion of conductance at low
bias flanked by peaks, is detected all over the scanned surface.
These features are still observed at the transition and persist on
cooling into the metallic phase. Polaron-binding-energy maps
reveal that polarons are not confined to regions embedded in a
highly-conducting matrix but are present over the whole field of
view both above and below the transition temperature.

\end{abstract}


%Use showkeys class option if keyword
                              %display desired
\maketitle

In manganite compounds such as \LCMOx\ ($0.2<x<0.5$), the coupling
between lattice, magnetism and transport leads to a transition
from metallic ($d\rho/dT>0$) to insulating ($d\rho/dT>0$) behavior
in tune with the suppression of ferromagnetic order at a
temperature \TMI\ \cite{Tokura00,Dagotto03}. Understanding the
transport properties of the insulating phase required the
consideration of the electron-phonon coupling mechanism
 in addition to double-exchange
interactions \cite{Millis95,Millis96}. In the strong coupling
limit electrons are bound by a surrounding lattice distortion
 forming polaronic quasiparticles \cite{Millis95,Millis96}. In the
insulating phase, polaron hopping is the dominant transport
mechanism and gives rise to the measured thermally-activated
resistivity \cite{Worledge96,Jaime96,DeTeresa97}. Although in
these context theoretical predictions \cite{Millis95,Millis96}
state that on cooling below \TMI\ spin order leads to electron
delocalization, evidence from numerous experimental techniques
\cite{Hundley95,Alexandrov01,Zhao00,Quijada98,Wei97,Ronnow06}
suggests the presence of polarons also at $T<T_{\rm{MI}}$. Optical
reflectivity \cite{Yoon98} and structural \cite{Louca97,Lanzara97}
investigations propose a crossover from a large to small-polaron
regime on warming across the MIT and a coexistence of both
polaronic phases at $T \sim T_{\rm{MI}}$. Therefore, the existence
of polarons in the metallic phase and the relevance of nanoscale
inhomogeneities \cite{Dagotto03} to the metal-to-insulator
transition (MIT) remain active subjects of debate. Insight into
this problem can be gained from scanning tunnelling spectroscopy
(STS) studies as the one presented here which probes local
electronic properties at the atomic scale.

\begin{figure}
\includegraphics[width=8.6cm]{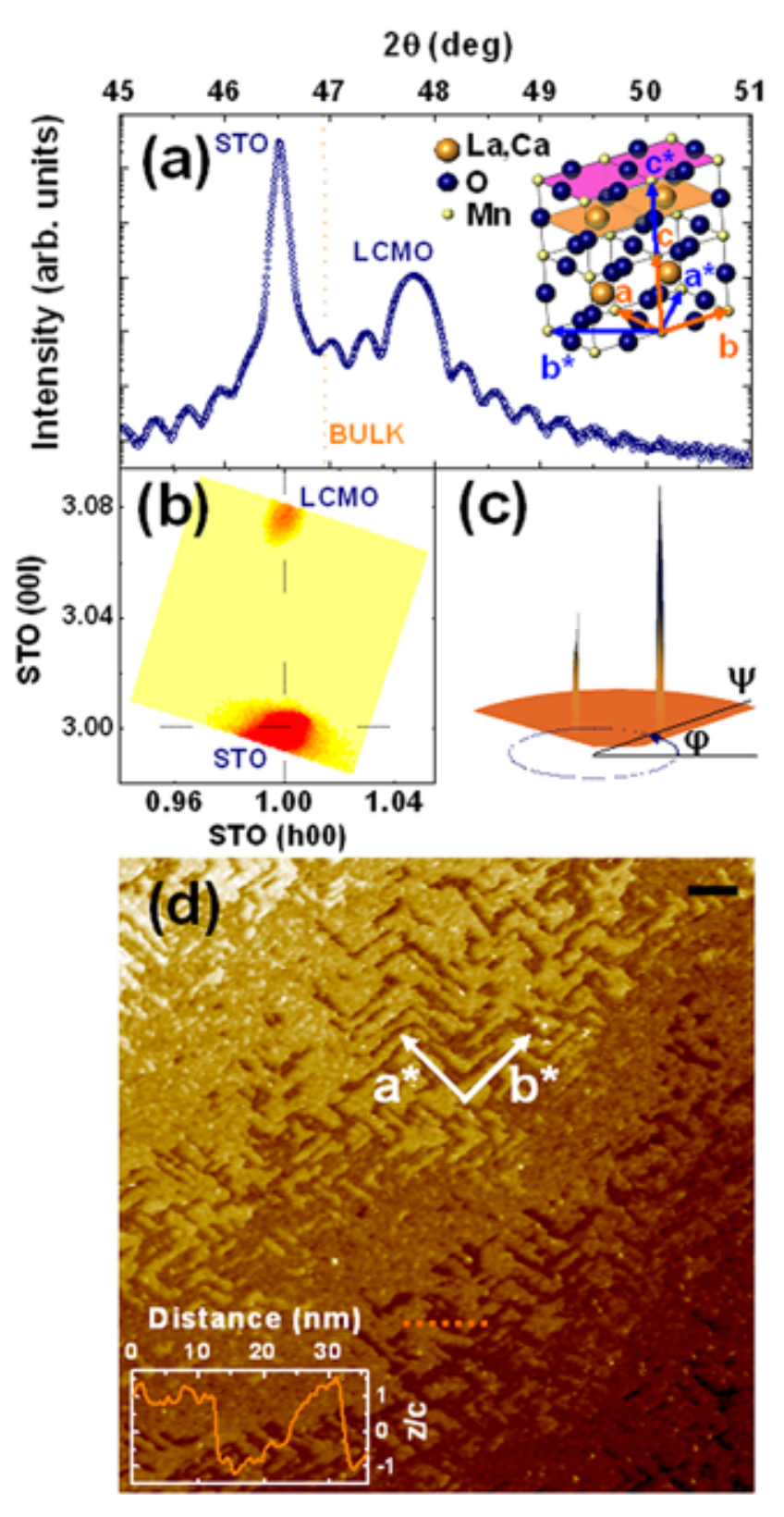}
\caption{\label{Figura1} (a) Main panel: X-ray diffractogram
around the (002) pseudocubic reflection and size-effect
oscillations. The film thickness was inferred to be 31\,nm and the
$c$ axis parameter (3.808$\pm$0.002)\,\AA. Insert: Schematic
crystalline structure. Pseudocubic ($\bf a$, $\bf b$, $\bf c$) and
orthorhombic ($\bf{a^*}$, $\bf{b^*}$, $\bf{c^*}$) lattice vectors
are indicated, as well as MnO$_2$ (pink) and (La,Ca)O (orange)
planes. (b) Reciprocal space map around the (103) reflection. (c)
Pole-figure measurement for the (103) direction. (d) Topograph
measured in the constant-current mode with regulation conditions
of 2\,V and 0.4\,nA at 144\,K. The $\bf{a^*}$ and $\bf{b^*}$
vectors of the film are at $45^{\circ}$ to the $\bf a$ and $\bf b$
axes of the substrate. The black scale bar represents 20\,nm. The
height profile in the insert corresponds to the dotted line
(orange).}
\end{figure}

\begin{figure}
\includegraphics[height=9cm,angle=-90]{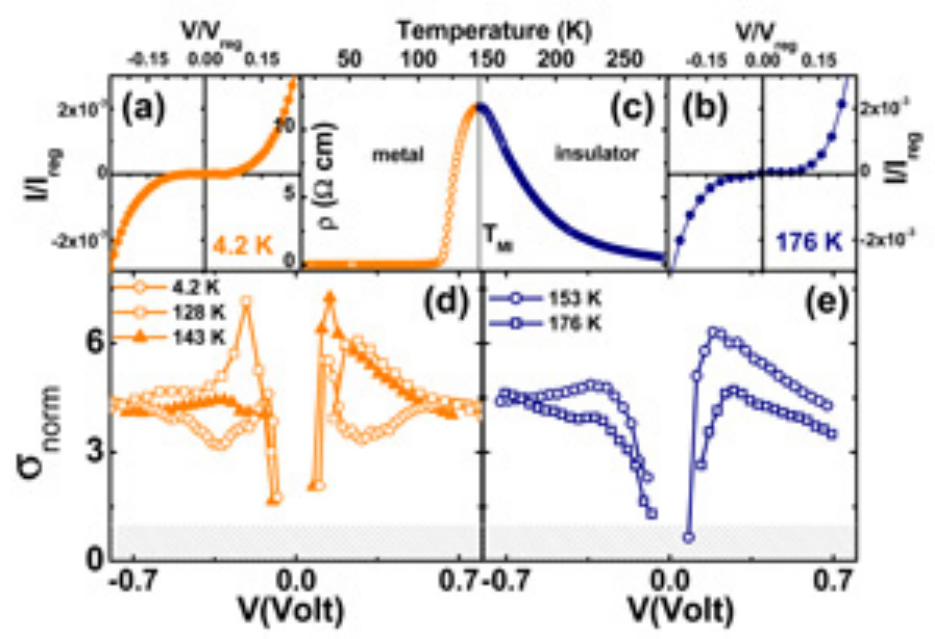}
\caption{\label{Figura2} Average tunneling $I(V)$ curves for (a)
4.2\,K (2\,V/0.5\,nA) and (b) 176\,K (1.3\,V/0.6\,nA).(c)
Macroscopic resistivity data: \TMI=143.9\,K is indicated by a
vertical line. (d) and (e) Selected normalized conductance curves
at temperatures below and above \TMI, respectively. When the
measured average current was below the experimental noise ($\sim
10^{-2}$\,pA) no estimation of $\sigma_{\rm{norm}}$ could be made
(hatched regions).}
\end{figure}

Although the phase separation scenario has gained popularity in
recent years, it is mostly supported by results from macroscopic
techniques \cite{Dagotto03}. Local spectroscopy studies in \LCMOx\
claim electronic phase separation
\cite{Faeth99,Becker97,Chen03,Moshnyaga06}, but a strong influence
of chemical and structural disorder present in the samples cannot
be ruled out. In contrast, a recent study in a fully-relaxed film
reports homogeneous conductance maps \cite{Mitra05}. However,
conductance maps at a fixed energy do not provide enough
information on the eventual presence, spectroscopic
characteristics and spatial distribution of polaronic
quasiparticles. Therefore, in this work we measure tunnel current
vs. voltage maps with nanometric spatial resolution in \LCMO\
(LCMO) films, tracking polaronic quasiparticles when cooling
across the MIT. The spectral signature of polarons is observed
both above and below \TMI\ all over the field of view.

The LCMO film studied in this work was grown on a SrTiO$_3$
substrate by RF sputtering at 675$^{\circ}$C, in 0.2\,Torr of a
50\%Ar-50\%O$_2$ mixture. Great effort was made to obtain a
single-crystalline fully-strained film. By controlling strain
\TMI\ can be reduced with respect to the bulk \cite{Song05}
allowing to access a wider temperature range of the insulating
phase. The structure of 3D perovskites does not present an easy
cleaving plane, so we thoroughly cleaned the surface of the film
with isopropanol in an ultrasonic bath. The sample was immediately
placed in the ultra-high vacuum chamber of a variable temperature
home-made STS system \cite{Kent92}, prior to cooling in $^4$He
exchange gas. This procedure allowed us to obtain reproducible and
high-quality topographic images as the one shown in Fig. 1.
Topographs and current vs. voltage ($I(V)$) maps were measured as
a function of temperature, with particular detail in the range
around \TMI. The tip, made of electrochemically etched Ir, was
grounded and the bias voltage $V$ was applied to the sample.

The thickness of the film, 31\,nm, was obtained from x-ray
reflectometry and confirmed by the Laue oscillations around the
(002) reflection, which is shown in Fig. 1 (a). The $c$ axis
parameter was found to be (3.808$\pm$0.002)\,\AA, 1.6\,\% smaller
than that of the bulk, as expected for tensile in-plane strain.
The reciprocal space map in Fig. 1(b) shows that the in-plane
lattice parameter matches that of SrTiO$_3$, i.e. the film is
fully strained. The (103) pole figure in Fig. 1(c) shows that the
film is single crystalline. The metal-to-insulator transition
temperature, $143.9\pm 0.3$\,K, was obtained from a four-point
resistivity measurement, see Fig. 2, as the temperature where
$d\rho/dT$ changes sign.  Topographs as the one in Fig. 1 (d)
reveal growth steps oriented along the $\bf{a^*}$ and $\bf{b^*}$
axis of the \LCMO\ orthorhombic structure. The height of steps is
systematically a multiple of $c$, the pseudocubic lattice
parameter, indicating the same terminating layer over the whole
surface. Whether it is MnO$_2$ or (La,Ca)O planes could not be
determined because, although rugosity was on average only 1\,\AA,
atomic resolution \cite{Ma05} was not achieved. As we show in a
previous report \cite{Seiro06}, local $I(V)$ curves are not
sensitive to the typical topographic features of our films and
rather homogeneous at the nanoscale. Therefore the spatial average
of $I(V)$ curves is representative of spectroscopic properties
over the whole sample surface.

In the STS experimental configuration, where a bias voltage V is
applied between the sample and the tip, the differential
conductance is given by
\begin{equation}
dI/dV \approx \frac{e^{2}}{\hbar} \rho_{s}(eV) \rho_{t}(0)
T(eV,z),
\end{equation}
\noindent where $\rho_{s}(eV)$ is the sample density of states,
the tip density of states $\rho_{t}(0)$ is assumed to be constant
and $z$ is the tip-sample separation \cite{Martensson89}. The
latter is set by the junction impedance $R_{\rm{T}}=
V_{\rm{reg}}/I_{\rm{reg}}$, that in STS experiments is typically
1\,G$\Omega$. If an electron is bound in a polaron, the process of
an electron tunnelling into the tip requires the unbinding of the
quasiparticle, with an energy cost of \Eb. As a consequence, STS
probes the local spectrum of the quasiparticles resulting from the
unbinding of polarons.

For bias voltages much smaller than the barrier height $\phi$, the
tunnel transmission coefficient $T$ does not depend on $V$ and the
measured differential conductance is proportional to
$\rho_{s}(eV)$. At voltages that are a non-negligible fraction of
$\phi$, as in the case of the measurements presented here, the
tunnel coefficient can no longer be considered
voltage-independent. For moderate voltages an adequate estimation
of $T(eV,z)$ is given by $I/V$ \cite{Martensson89}. Therefore, the
numerically derived $I(V)$ curves were normalized to obtain
$\sigma_{\rm {norm}}=(dI/dV)/(I/V) \propto \rho_{s}(eV)$, the
normalized conductance. At $V\sim 0$, the noise on the tunnel
current hinders an appropriate estimation of $T(eV,z)$ with $I/V$.
As a consequence, at low voltages, $T(eV,z)$ was estimated with a
WKB-like tunnel transmission coefficient \cite{Ukraintsev96}.

In this work 10$^4$ single $I(V)$ curves were spatially averaged
over 60$\times$60\,nm$^2$. Examples of average $I(V)$ curves in
the metallic and insulating phases are shown in Fig. 2. At all
measured temperatures $dI/dV$ at zero bias is below the
experimental resolution of typically $10^{-4}$\,nS. In the
high-temperature phase $\sigma_{\rm{norm}}$ presents a depletion
around the Fermi level flanked by conductance peaks, as shown in
Fig. 2 (c). These features are the spectral signature of polarons
and the half-distance between the peaks, $\Delta_{\rm B} $, is a
measure of the polaron binding energy \cite{Wei97}. At 176\,K,
well above \TMI, $\Delta_{\rm B} =(0.30\pm 0.07)$\, eV is
consistent with the polaron binding energy obtained by fitting the
high-temperature resistivity data
 with an adiabatic small-polaron model, see Fig. 3. Within this model $\rho = \rho_0 T
\exp(E_A/kT)$ \cite{Worledge96} and the binding energy can be
estimated as $E_{\rm b} \approx 2 E_A  = (0.24 \pm 0.02)$\,eV
\cite{Worledge96}.

In the metallic phase these polaronic characteristics persist even
down to 4.2\,K, see Fig. 2 (d). Although in contrast to the
theoretical prediction that polarons disappear below \TMI\
\cite{Millis95,Millis96},  our local spectroscopic results are
consistent with data from different experimental techniques
\cite{Hundley95,Alexandrov01,Zhao00,Quijada98,Wei97,Hartinger04}.
On decreasing temperature $\Delta_{\rm B}$ globally shifts to
lower energies, as shown in Fig. 3. This behavior is consistent
with the temperature evolution of polaron binding energies
obtained from macroscopic optical reflectivity data
\cite{Hartinger04}. In spite of all these evidence, the role of
polarons in the transport properties of the metallic phase is
currently under discussion \cite{Mannella05,Zhao00b}.

Remarkably, on top of the global temperature behavior,
$\Delta_{\rm B}$ presents a dip at \TMI. Although STS probes the
quasiparticle spectrum at the surface, the extreme sensitivity of
$\Delta_{\rm B}$ to the MIT and its quantitative agreement with
the \Eb\ estimated from transport data indicate that the
$\sigma_{\rm{norm}}$ measured in this work is representative of
bulk properties.

Since normalized conductance curves in Fig. 2 come from a spatial
average of $10^4$ curves, it could be argued that the polaronic
signal observed below \TMI\ comes from remnant domains of the
high-temperature phase. Polaron-binding-energy maps reveal that
this is not the case: The spectroscopic signature of polarons is
detected over the whole field of view of $60\times 60$\,nm$^2$
with a spatial resolution of $\sim 2$ unit cells. The
binding-energy distributions are Gaussian with a dispersion of up
to 20\% which does not significantly change with temperature (it
varies only 2.5\,\% from 176 to 140\,K). No bimodal distribution
of $\Delta_{\rm B}$ is detected within 0.03\,eV, a resolution
which is of the order of the thermal energy.

\begin{figure}
\includegraphics[height=8.6cm,angle=-90]{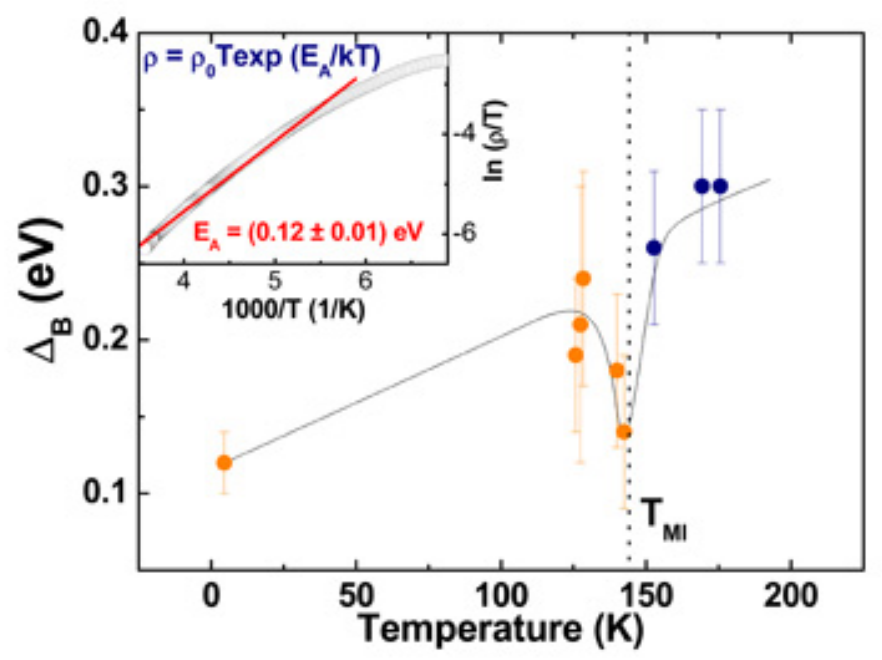}
\caption{\label{Figura3}Temperature dependence of the polaron
binding energy estimated from normalized conductance curves as
half the peak-to-peak separation. The dotted line is a guide to
the eye. Insert: Fit (red line) of the high-temperature
resistivity data (open symbols) with a small-polaron adiabatic
model (see text).}
\end{figure}

\begin{figure}
\includegraphics[width=8.6cm]{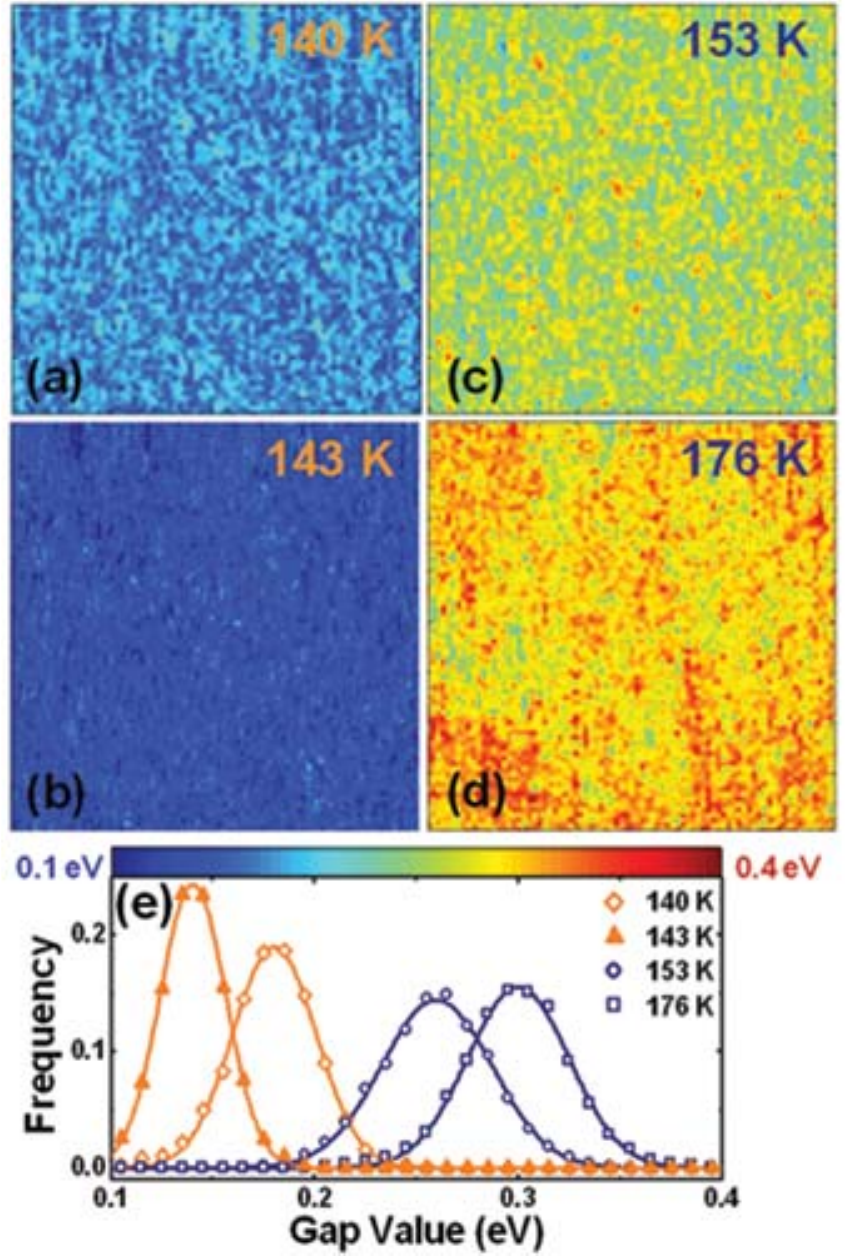}
\caption{\label{Figura4}Mapping of local $\Delta_{\rm B}$ values
at (a) 140, (b) 143, (c) 153 and (d) 176\,K. All maps correspond
to 60$\times$60\,nm$^2$ areas with a resolution of 7.5\,\AA/pixel.
(e)  Distributions of $\Delta_{\rm B}$ values (symbols) obtained
from the maps fitted with a Gaussian law (lines).
   The average noise of local I(V) curves in all maps
    is comparable, ranging from 0.2 to 0.4\,pA.}
\end{figure}

At low temperatures, the non-linear dependence of $I$ on the bias
voltage might seem in contradiction with a metallic state. This
striking result can have different origins. Firstly, in order to
probe polarons spectroscopy has to be performed in the energy
scale of \Eb, which is a non-negligible fraction of $\phi$. As a
consequence, the tip-sample separation is large and
simultaneously, the transmission coefficient increases with
voltage, resulting in a non-linear $I(V)$ characteristic.
Secondly, the macroscopic resistivity of the 31\,nm film has a
value of 4\,m$\Omega$~cm at 4.2\,K, 3 orders of magnitude larger
than for a common metal, implying that the number of carriers
and/or their mobility is low. Recent angle-resolved photoemission
spectroscopy (ARPES) measurements in the bilayer manganite
La$_{1.4}$Sr$_{1.6}$Mn$_2$O$_7$, revealed the presence of
``nodal'' quasiparticles that would account for the poor
conductivity of the metallic phase \cite{Mannella05}. Since the
measured tunnel current comes from a reciprocal-space integration
of the single particle spectral function (as detected by ARPES)
\cite{Fischer06}, the contribution to the tunnel conductance at
$E_F$ will be extremely small. Therefore, a low tunnel conductance
at zero bias is not at odds with metallic behavior. In addition to
the measurements presented in this work, previous STS studies
report highly nonlinear spectra in the metallic phase of a
La$_{0.7}$Ca$_{0.3}$MnO$_3$ film at 77\,K \cite{Wei97} and in a
cleaved La$_{1.4}$Sr$_{1.6}$Mn$_2$O$_7$ crystal \cite{Ronnow06}.
In order to study the spectral features close to the Fermi level,
high energy-resolution measurements at small $z$ are currently
underway and will be the subject of a future report. We stress
that this work focuses on the temperature evolution and spatial
homogeneity of polaronic spectral features that are manifest at an
energy scale of $\sim 0.1-1$\, eV.

In summary, we present tunneling spectroscopic evidence for the
presence of polarons in the insulating phase and their persistence
on cooling through the insulator-to-metal transition down to
4.2\,K.  The spectroscopic signature of polarons is detected over
the whole field of view at all measured temperatures. This novel
result brings forth the challenge of understanding the role played
by polaronic quasiparticles in the metallic phase of manganites.

\begin{acknowledgments}
This work was supported by the Swiss National Science Foundation
and the MaNEP program of the Swiss National Center of Competence
in Research.
\end{acknowledgments}

\end{document}